\documentclass{article}
\usepackage[ansinew]{inputenc}
\usepackage{times}
\usepackage[T1]{fontenc}
\usepackage{graphicx}
\usepackage{geometry}
\usepackage{amssymb}
\usepackage{amsmath}

\usepackage{rotating}

\usepackage{tcolorbox}
\usepackage{xcolor}
\colorlet{shadecolor}{gray!25}
\usepackage{subfig}
\usepackage{float}

\title{Tests for strict monotonic trend in bio-medical dose-response relationships (respective concentration-response or exposure-response relationships)
	- a biostatistical perspective}
\author{\normalsize Ludwig A. Hothorn\\ 
\footnotesize Im Grund 12, D-31867 Lauenau, Germany (e-mail:ludwig@hothorn.de)\\ \scriptsize(retired from Leibniz University Hannover)}

\begin{document}

\maketitle
\begin{abstract}
Evidence of a global trend in dose-response dependencies is commonly used in bio-medicine and epidemiology, especially because this represents a causality criterion.
However, conventional trend tests indicate a significant trend even when dependence is in the opposite direction for low doses when the high dose alone has a superior effect. Here we present a trend test for a strictly monotonic increasing (or decreasing) trend, evaluate selected sample data for it, and provide corresponding R code using CRAN packages. 
\end{abstract}

\section{The problem} \label{problem}
The proof of a trend in selected data is extensively demanded in guidelines, e.g. for mutagenicity bioassay \cite{OECD2015}, and carried out in publications of different disciplines, e.g. in genetic association studies \cite{manning2023},  in long-term carcinogenicity bioassay \cite{hothorn2022robust} or in clinical dose finding trials \cite{bretz2021}. Notice, common-used time trends are analyzed particularly, e.g. for proportions over different years by the Armitage  trend test \cite{rezaei2023}. This is however biased, since the times, e.g. years, are not independent, as required for the quantitative covariate of the Armitage test \cite{Armitage1955}. In this article, a completely randomized one-way design is considered, on the one hand for dose as a qualitative factor, e.g. with the levels 'control, low, medium, high' or on the other hand for a quantitative covariate, e.g. concentration: 0,10,100, 500 mg. In contrast to observational epidemiological exposure studies we assume here discrete chosen concentration levels. Often the randomly distributed exposures are categorized post-hoc \cite{mabikwa2017}, e.g. by quintiles, which can be problematic \cite{greenland1995}.  I.e. we consider here multiple contrast tests for a qualitative factor and regression models for a quantitative covariate. \\
Why is the detection of a trend so specific compared to general inference? A possible explanation is the evidence of a trend as a causation criterion. For example, Hill wrote that \textit{'if a dose response is seen, it is more likely that the association is causal.'} (criteria 5) \cite{steenland2004} (an recent discussion \cite{ioannidis2016}, considering a biological gradient \cite{fedak2015}). \\

A good sample description is provided by the quote from \cite{steenland2004}: \textit{In traditional epidemiology, a monotonic biological gradient, wherein increased exposure resulted in increased incidence of disease, provides the clearest evidence of a causal relationship. However, Hill acknowledged that more complex dose-response relationships may exist, and modern studies have confirmed that a monotonic dose-response curve is an overly simplistic representation of most causal relationships. In fact, most dose-response curves are non-linear and can even vary in shape from one study to the next 'AND molecular changes within the no-observable-adverse-effect level (NOAEL) may not contribute to disease and are more indicative of a threshold dose response}. I.e.  the shape of dose-response is not an assumption neither for a contrast test nor a regression model, it is a data-dependent outcome. On the one hand, versatile trend tests should be sensitive against many shapes, whereas sensitive against all possible shapes (including partial non-monotonic) is an illusion. On the other hand most contrast test or regression model-based test suffers to be significant as long as $\mu_0-\mu_k$ is large enough, irrespective of the shape of the entire dose-response relationship, including non-monotonic reverse effects at lower doses. In the following, a specific variant of trend tests is presented, which show the detection of strictly monotonic trends only, i.e. at low doses there should be no effect reversal, but non-significant effects are allowed. Three non-monotonic examples will explain this issue: i) left panel: number of young per adult in daphnia whole
effluent toxicity assay \cite{Hothorn2016} (available in library(SiTuR)), ii) mid panel: proportions of thyroid tumors in rats treated with several concentrations of ethylene thiourea \cite{graham1975} (available in library(EnvStat), iii) right panel: summarized proportions of breast tumors exposed to categorized increasing concentrations of alcohol in a case-control study \cite{rohan1988} (available in library(dosresmeta)).

\begin{figure}
	\centering
		\includegraphics[width=0.3\textwidth]{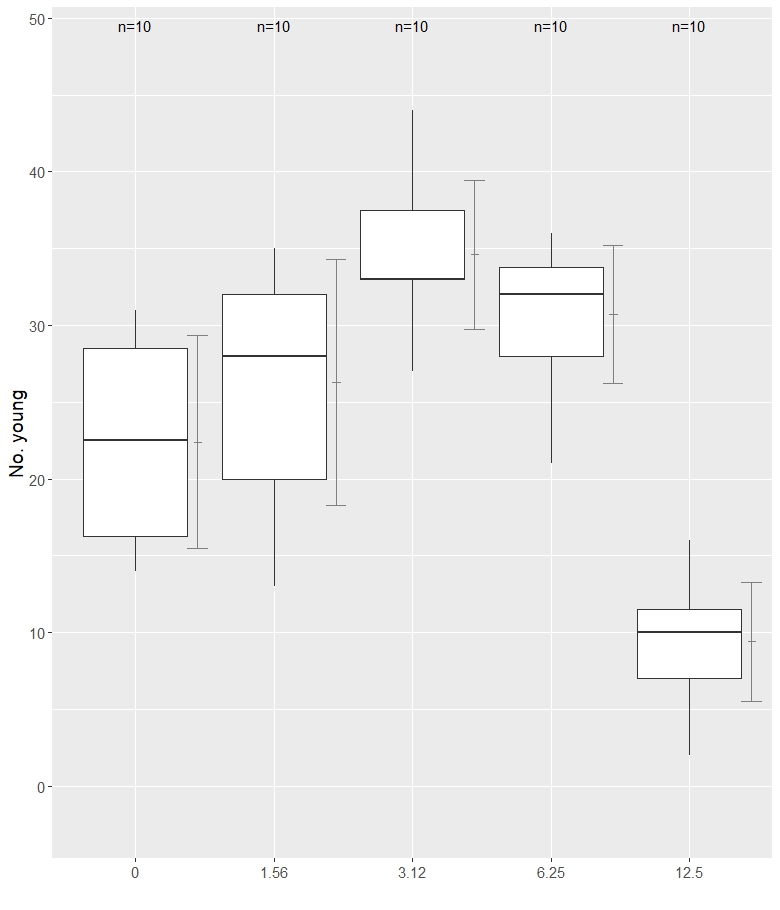}
		\includegraphics[width=0.3\textwidth]{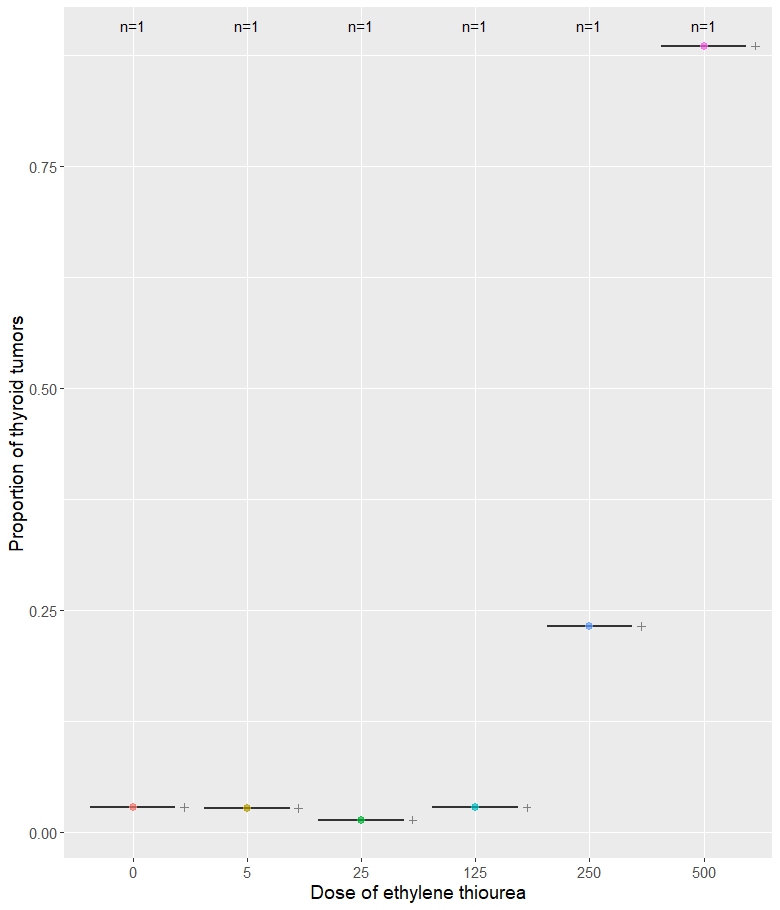}
		\includegraphics[width=0.3\textwidth]{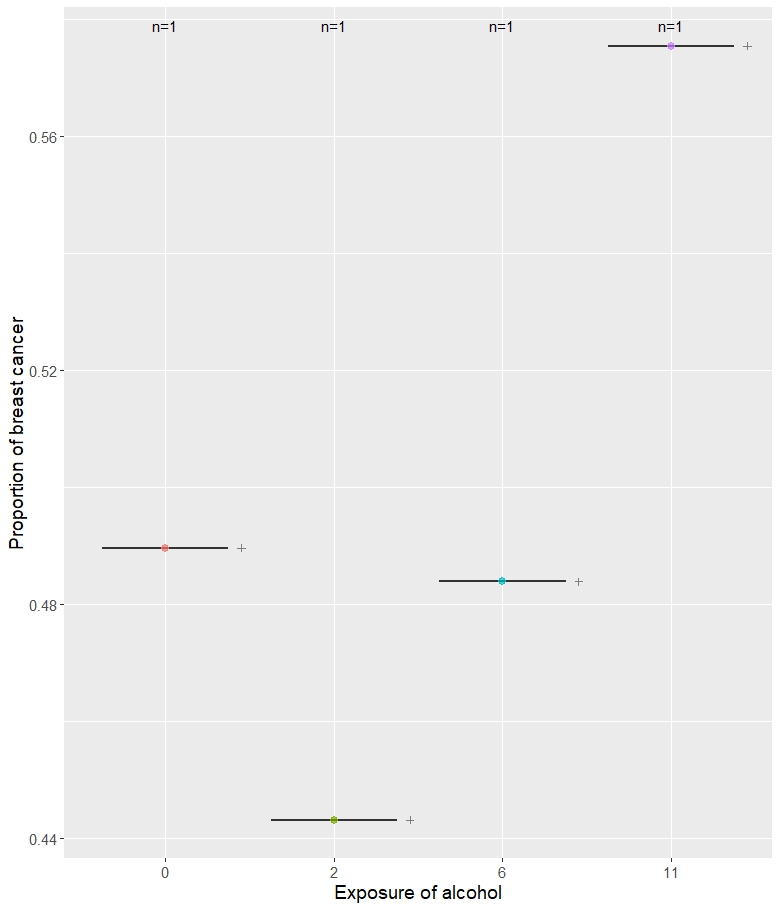}
	\caption{Three data examples for non-monotonic trend}
	\label{fig:Daphnia}
\end{figure}

These examples reveal certain non-monotonicity at low doses or concentrations. When assuming a monotonic model, the question arises, what is more important: this model assumption, even if the data contradict it, or special models which consider exactly such non-monotonicities. Here the latter approach is taken. The context can be complex. Therefore, we consider exactly one study, i.e., no meta-analysis or mixed models across multiple dose-response dependencies. Likewise, we restricted ourselves to the trend statement per se, i.e., estimation of relevant doses, such as the minimum effective dose (MED), are not considered here.

%%%%%%%%%%%%%%%%%%%%%%%%%%%%%%%%%%%%%%%%%%
\section{Tests for strict monotonic trend} \label{main}

\subsection{Different formulations of alternative hypotheses} \label{hypo}
To non-statisticians it seems to be clear what a trend is: increasing effects with increasing dose levels. From a statistical perspective it is not. The common trend tests, such as Williams \cite{Williams1971}, Jonckheere \cite{Neuhauser1998}, Armitage \cite{Armitage1955}, Tukey \cite{tukey1985, Schaarschmidt2021} are significant if $H_0:\mu_k-\mu_0$ is strong enough in $H_1$, irrespective of the lower level inferences- up to opposite effect sides. A simulated data example visualize this issue: both Williams contrast test and Tukey regression tests reveal a highly significant increasing trend ($p<0.0001$), but researchers may have difficulty inferring an increasing trend from these data.

\begin{figure}[htbp]
	\centering
		\includegraphics[width=0.4\textwidth]{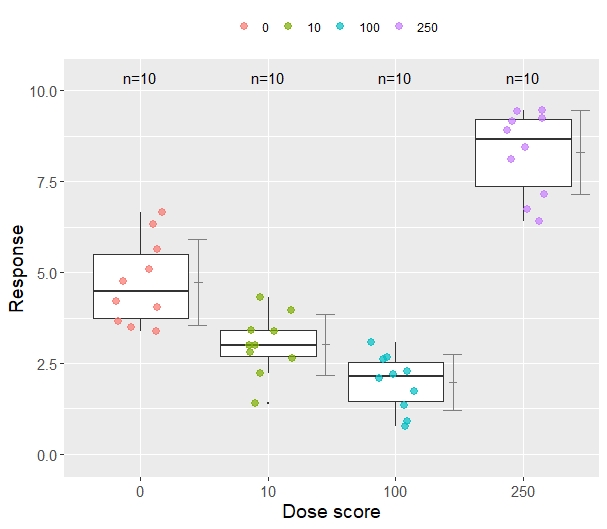}
	\caption{Simulated increasing, but non-monotonic, dose-response relationship}
	\label{fig:SimEx}
\end{figure}

In the following, multiple contrast tests with monotonic ordered alternative are used to derive appropriate tests. 
The dilemma lies in the hypothesis space of one-sided, ordered hypotheses for a messily written $H_0$:\\
$$H_0:\mu_0=\mu_1=,...,=\mu_k$$ \\
$$H_1:\mu_0\leq \mu_1 \leq,...,\leq \mu_k| \mu_0<\mu_k$$ \\

The complexity of ordered $H_1$ become clear, when its is decomposed into the underlying elementary hypotheses.
As an example a design with k=3+1 is used for simplicity whereas the global alternative can be decomposed into $7$ elementary order-restricted alternatives:
\[ H_1^{\xi=1,a}:\mu_0 < \mu_1= \mu_2= \mu_3\]
\[ H_1^{\xi=1,b}:\mu_0 < \mu_1< \mu_2= \mu_3\]
\[ H_1^{\xi=1,c}:\mu_0 < \mu_1= \mu_2< \mu_3\]
\[ H_1^{\xi=1,d}:\mu_0 < \mu_1< \mu_2< \mu_3\]
\[ H_1^{\xi=2,a}:\mu_0 = \mu_1< \mu_2= \mu_3\]
\[ H_1^{\xi=2,b}:\mu_0 = \mu_1< \mu_2< \mu_3\]
\[ H_1^{\xi=3,a}:\mu_0 = \mu_1 = \mu_2< \mu_3\]
All elementary hypotheses represent elementary trends. Notice, some scientists consider $H_1^{\xi=1,a}$ and/or $H_1^{\xi=3,a}$ not as trend- related procedure are described in the Appendix.

%%%%%%%%%%%%%%%%%%%%%%%%%%%%%%%%%%%%%%%%%%%%%
\subsection{Strict monotonic alternatives} \label{shyp}
In principle, trend test can be formulated for two-sided hypotheses- here we consider the common case of one-sided hypotheses, i.e. either and increase or and decrease separately. In the following, three types of trend test will be formulated: i) the common used, ii) the absolute monotonic trend, where ALL  $\mu_i-\mu_0;  i=k,(k-1),...,1 $ must be under $H_1$, iii) strict monotonic trend, where the  $\mu_i-\mu_0;  i=k,(k-1),..,j$ follows a trend claim AND $\mu_i-\mu_0;  i=j, (j-1),...,1$ must be at least non-inferior. I.e. we tolerate a small $\eta$ change in the opposite direction.\\
For case i) the literature is overwhelming, e.g. for dose as factor the Williams trend test \cite{Williams1971} or for dose as covariate the Tukey trend test \cite {tukey1985}. Further rather specific approaches are iv) a trend where each change point contrast contribute- see chapter \ref{change} in the Appendix and v) a trend where each dose increment contribute - see the Appendix.\\
For case ii), denoted as global absolute trend, a test is under $H_1$ if all pairwise contrasts for $\mu_i-\mu_0$, i.e. Dunnett-type contrasts are significant, each at level $\alpha$- a special version of an intersection-union-test \cite{hothorn2021}, see chapter \ref{mono}.\\
For case iii), denoted as strict monotonic trend,  let us assume a particular one-sided alternative $$H_{1'}:\mu_0= \mu_1 =,...,= \mu_{k-1}<\mu_k)$$ precisely than is $$H_{1'}:\mu_0\geq\mu_1\geq,...,\geq\mu_{k-1}<\mu_k)$$, i.e. decreasing effects at lower doses $j<k$ are possible within this framework. This approach combines point-zero-$H_0$-tests on superiority for at least $$H_{1''}:\mu_0= <\mu_k $$ and tests on non-inferiority for at least  $$H_{1'}:\mu_0\geq\mu_1+\eta$$.At least means it can be more than just $\mu_k$ superior to control and/or more than just $\mu_1$ non-inferior to control.

%%%%%%%%%%%%%%%%%%%%%%%%%%%%%%%%%%%%%%%%%%%%%%%%%%%%

\subsection{A proposal for an approach, able to claim a global absolute monotonic trend}\label{abs}
The specific alternative $H_1: \mu_0<µ_1<,...,<\mu_k$ does not guarantee the claim for an absolute trend, simple because ordered $\mu_i$ do not necessarily generate ordered p-values $p_i$.  A specific shortcut of the closed testing procedure (CTP) \cite{marcus1976} can be used \cite{Hothorn2021a}:
 $(\mu_0<µ_1<,...,<\mu_k) \Rightarrow (\mu_0<µ_1<,...,<\mu_{k-1}) \Rightarrow...\Rightarrow (\mu_0<µ_1)$. Each of these sub-hypotheses can be tested with simple pairwise comparisons  $\mu_0-\mu_i; i=k, k-1,...,1$ because the condition $\mu_0<\mu_i; i=k, k-1,...,1$ alone is sufficient to claim a trend. Either pairwise tests (e.g. Welch-t-tests) can be used, or better pairwise contrasts for designs with small sample sizes.\\
A perfect example are Shirley s reaction time data:
\begin{figure}[htbp]
	\centering
		\includegraphics[width=0.35\textwidth]{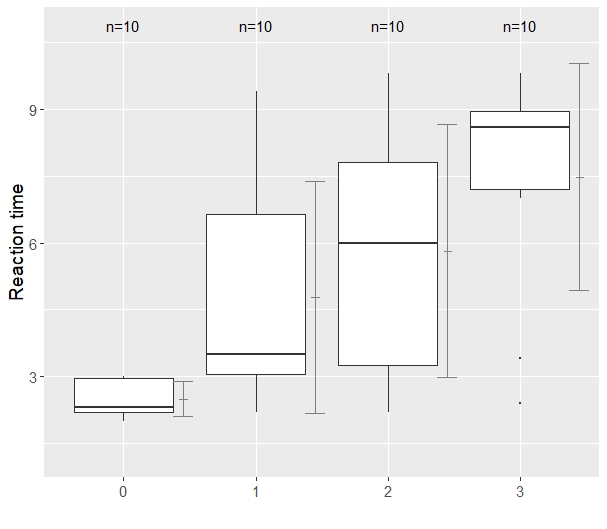}
	\caption{Reaction time assay}
	\label{fig:Shir}
\end{figure}

\footnotesize
\begin{verbatim}
data("reaction", package="nparcomp")
reaction$Dose <-as.factor(reaction$Group)
mod2<-lm(Time~Dose, data=reaction)
absglobal<-summary(glht(mod2, linfct = mcp(Dose="Dunnett"), 
           alternative="greater", vcov = sandwich, p.adjust="none"))
\end{verbatim}
\normalsize

\begin{table}[ht]
\centering\footnotesize
\begin{tabular}{rrrrr}
  \hline
Contrast & Estimate & SE & test statistics & adjusted p-value \\ 
  \hline
1 - 0 &  2.2800 & 0.7913 & 2.8813 & 0.0099 \\ 
  2 - 0 & 3.3200 & 0.8620 & 3.8517 & 0.0007 \\ 
  3 - 0 & 4.9800 & 0.7743 & 6.4315 & $<0.00001$ \\ 
   \hline
\end{tabular}
\caption{Reaction time assay data: global absolute monotonic trend}
	\label{tab:MShir}
\end{table}

Since all unadjusted p-values are significant, an absolute trend can be claimed since $p_{3-0}<p_{2-0}<p_{1-0}$ and all p-values are significant:

%%%%%%%%%%%%%%%%%%%%%%%%%%%%%%%%%%%%%%%%%%%%%%%%%%%%%%%%%%%%%%%%%%%%%%%%%%%%%%%
\section{A proposal for an approach, able to claim a strict monotonic trend} \label{mono}
Instead of p-values, confidence intervals are recommended: additional to the significant/non-significant decision (as p-values), they provide directional decision and relevance related effect sizes with their amount of uncertainty. Because no uni-directional tests are considered in this approach, two-sided confidence intervals are used. Equivalence tests can be performed as TOST (two one-sided tests) and translated into compatible $(1-2\alpha)=0.90$ confidence intervals. These confidence intervals can be used also as one-sided $(1-\alpha)$ limits for trend hypotheses.\\
Two families of hypotheses are combined: 1) the union-intersection-type hypotheses for trend within the Williams trend test and 2) the marginal elementary hypotheses $H_0^{0i}$. Commonly, the elementary hypotheses are formulated as point-zero hypotheses, i.e. $H_0^{0i}: \mu_0=\mu_i  vs. H_1^{0i}: \mu_0<\mu_i$. Here we modify into non-inferiority hypotheses $H_0^{0i noninf}: \mu_0\geq\mu_i-\eta  vs. H_1^{0i noninf}: \mu_0<\mu_3-\eta$. Translated into confidence intervals: the upper limit should by smaller the threshold parameter $\eta$ (in the case of increasing trend; the lower limits for decreasing trend). The both families of hypotheses are combined by means of an intersection-union test principle \cite{hothorn2021} presented as two-sided $(1-2\alpha)$ confidence limits: a strict monotonic trend exists when at least one of the simultaneous Williams-type confidence limit is larger/less the the point of $H_0$ of zero, \textbf{AND all} elementary hypotheses are at least non-inferior, i.e. less/larger than $\eta$. Unfortunately, the margin of non-inferiority is rarely available a-priori. Therefore, the interpretation for strict monotonic trend is performed data-dependent, post-hoc, i.e. if the maximum/minimum lower/upper limit is still acceptable as not in the undesired direction (see the interpretation of the example below).\\
Notice, the above approach's outcome is an $\eta$-dependent test, where $\eta$ is rarely defined a-priori.  Here, the empirical, post-hoc comparison between the with of superiority intervals (Williams trend test contrasts) and the non-inferiority intervals (pairwise contrasts against control), see the related discussion \cite{campbell2018}. This is certainly a weak feature of the approach, but probably inherent, just as with most non-inferiority tests in practice.

%%%%%%%%%%%%%%%%%%%%%%%%%%%%%%%%%%%%%%%%%%%%%%
\subsection{Evaluation of the daphnia data} \label{daphnia}
The daphnia aquatic toxicity example data are used \cite{Hothorn2016} where the decreasing trend is interrupted by increasing low dose effects:
\begin{figure}[htbp]
	\centering
		\includegraphics[width=0.45\textwidth]{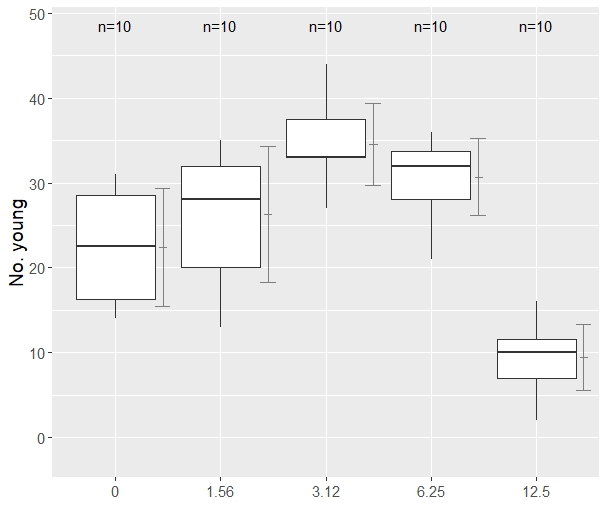}
		\caption{Daphnia data: box plots}
	\label{fig:Daph}
\end{figure}
Technically, the 2-sided simultaneous $0.90$ confidence interval for the unadjusted, elementary pairwise contrasts are estimated by the Dunnett test for a level of 4*0.10 (4 ... anti-Bonferroni for 4 contrasts). Instead of the common used MQR variance estimator, the sandwich estimator is used, to be robust against variance heterogeneity \cite{Hasler2008, Herberich2010}.

\footnotesize
\begin{verbatim}
library(multcomp)
globalW<-confint(glht(mod1, linfct = mcp(Conc="Williams"), vcov = sandwich),level=0.90)
levelNew=(1-0.10*4)
globalDu<-confint(glht(mod1, linfct = mcp(Conc="Dunnett"), vcov = sandwich),level=levelNew)
\end{verbatim}
\normalsize
\begin{figure}
	\centering
		\includegraphics[width=0.4\textwidth]{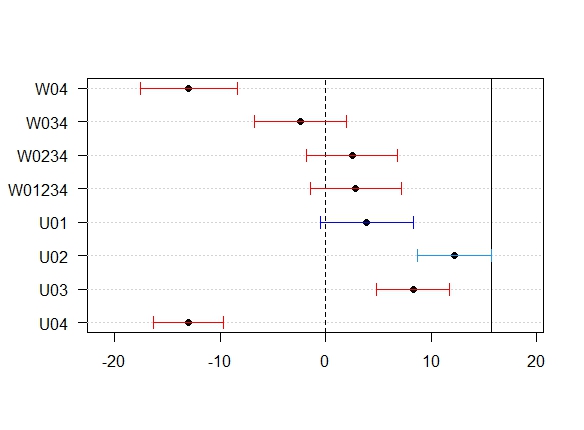}
	\caption{Daphnia data example: estimated intervals (W ... Williams, U...unadjusted vs. control)}
	\label{fig:DUWI}
\end{figure}

The common used Williams test claim for trend (at least one, anyone contrast is under $H_1$), since the upper limit for $\mu_0-\mu_{12.5}$ is less the zero (a clear loss of more than 8 number of young). A claim for strict trend is possible only, if an increase of more than 8 young (for $\mu_0-\mu_{3.12}$  is still tolerable as decreasing- a rather unlikely interpretation, see Figure \ref{fig:DUWI}.\\

%?????Since the tolerance threshold is rarely available, a crude empirical approach (without statistical properties): the one-sided p-values for the elementary tests must be $<0.5$, i.e. not in the opposite direction.

%%%%%%%%%%%%%%%%%%%%%%%%%%%%%%%%%%%%%%%%%%%%%%%%%%%%%
\subsection{Evaluation of the Shirley's reaction time data} \label{shir}
Since Shirley’s reaction time data already show an absolute monotonic trend (see chapter \ref{abs}, they should also show a strict monotonic trend, since this is hierarchically implied.
%\begin{figure}[htbp]
%	\centering
%		\includegraphics[width=0.470\textwidth]{D:/PUB/Zurich2023Trend/SH.jpeg}
%	\caption{Shirley reaction time data: box plots}
%	\label{fig:SH}
%\end{figure}

\begin{figure}[htbp]
	\centering
		\includegraphics[width=0.4\textwidth]{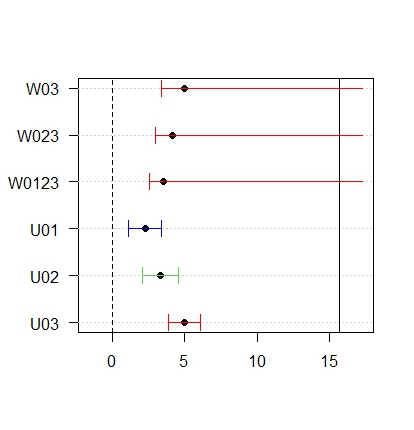}
	\caption{Shirley reaction time data: estimated intervals (W ... Williams, U...unadjusted vs. control)}
	\label{fig:SH1}
\end{figure}

The decision scheme of confidence intervals reveals i) a global trend exists since all lower limits are greater 0 for the Williams-type contrasts and ii) a strict monotone trend because additionally all lower limits for many-to-one comparisons are even greater than zero (non-inferiority is hierarchical implied into superiority tests, see Figure \ref{fig:SH1}.

%%%%%%%%%%%%%%%%%%%%%%%%%%%%%%%%%%%%%%%%%%%%%%%%%%%
\subsection{Evaluation of case-control data} \label{green}
\begin{figure}
	\centering
		\includegraphics[width=0.4\textwidth]{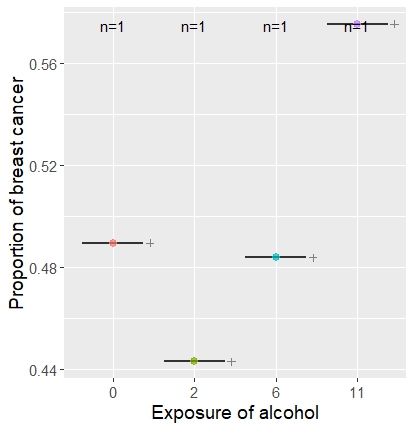}
	\caption{Case-control data for breast cancer and alcohol intake}
	\label{fig:Crippa}
\end{figure}

A data-set was selected which contains the summarized dose-response results from a case-control study on alcohol and breast cancer  \cite{Greenland1992}. These data were analysed by a log-linear model \cite{crippa2016} and with a clear recommendation, however based on strict monotonicity, see chapter \ref{quant}. The data are however no-monotonic, i.e. only the highest exposure supports an assumed increasing trend, see Figure \ref{fig:Crippa}.

A standard generalized linear model (GLM) with log-odds ratios as effect size is used for the qualitative factor levels for alcohol exposure.

\footnotesize
\begin{verbatim}
library(toxbox)
library(dosresmeta)
data(cc_ex)
cc_ex$prop<-cc_ex$case/cc_ex$n
cc_ex$Dose<-as.factor(cc_ex$dose)
library(multcomp)
mod4 <- glm(cbind(case,control)~Dose,data=cc_ex, family=binomial(link="logit"))
CIOR<-confint(glht(mod4, linfct = mcp(Dose = "Williams")), level=0.90)
levelNew3=(1-0.10*3)
globalDun<-confint(glht(mod4, linfct = mcp(Dose="Dunnett")),level=levelNew3)
\end{verbatim}

\begin{figure}
	\centering
		\includegraphics[width=0.4\textwidth]{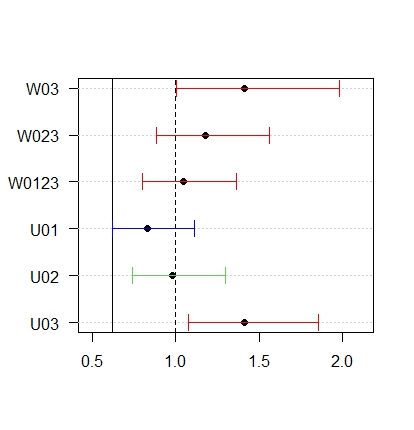}
	\caption{Confidence limits for odds ratios of the case control data (W ... Williams, U...unadjusted vs. control)}
	\label{fig:Green1}
\end{figure}

\normalsize
The confidence intervals for the transformed odds ratios reveal an increasing trend by means of the Williams contrasts, clearly only for contrast $p_3-p_0$. Because two of the three pairwise contrasts against control reveal lower limits less than 1 and for the contrast $p_2-p_0$ even a rather small value of 0.62 was estimated, it is hard to believe in a strict trend, see Figure \ref{fig:Green1}.

%%%%%%%%%%%%%%%%%%%%%%%%%%%%%%%%%%%%%%%%%%%%%%%%%%%%%%%%%%%%%%%%%%%%%%%%%%%%%%%%%%%%%
\section{A proposal for an approach, able to claim a strict monotonic trend: dose modeled quantitatively}\label{quant}
\begin{figure}
	\centering
		\includegraphics[width=0.4\textwidth]{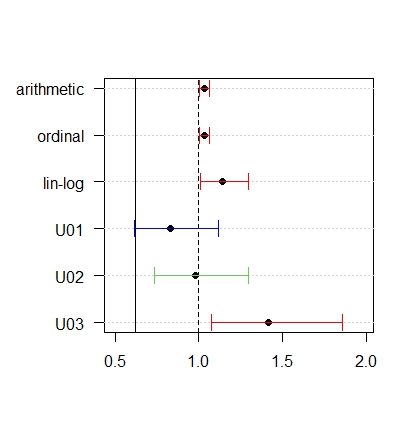}
	\caption{Confidence limits for odds ratio: dose as covariate (U...unadjusted vs. control) }
	\label{fig:GreenTukey}
\end{figure}

\normalsize
The above approach is modified for regression-type models by replacing the Williams trend test by the  Tukey-type tests \cite{Schaarschmidt2021}.\\
The above example with a dose-response relationship in a summarized case-control study on alcohol and breast cancer is used again. Analysed by \cite{crippa2016} by a log-linear model, the slope for relative risk linear model was estimated to 0.045 (one-sided p=0.028) concluding 'on the exponential scale, every 1 gram/day increase of alcohol consumption was associated with a $4.6 \% (exp(0.0454) = 1.046)$ higher breast cancer'. But the dose-response pattern is non-monotonic! In our opinion, this conclusion is biased. \\
\footnotesize
\begin{verbatim}

Dlevelw<-as.numeric(levels(cc_ex$Dose))
S0W<-log(Dlevelw[2])- log(Dlevelw[3]/Dlevelw[2])*(Dlevelw[2]-Dlevelw[1])/(Dlevelw[3]-Dlevelw[2])
cc_ex$DoseN<-as.numeric(as.character(cc_ex$Dose))
cc_ex$DoseO<-as.numeric(cc_ex$dose)
cc_ex$DoseL<-log(cc_ex$DoseN)
cc_ex$DoseLL<-cc_ex$DoseL
cc_ex$DoseLL[cc_ex$DoseN==Dlevelw[1]] <-S0W

pN <-glm(cbind(case,control)~DoseN,data=cc_ex, family=binomial(link="logit"))
pO <-glm(cbind(case,control)~DoseO,data=cc_ex, family=binomial(link="logit"))
pLL <-glm(cbind(case,control)~DoseLL,data=cc_ex, family=binomial(link="logit"))

gre <- glht(mmm(covarMa=pN, ordinMa=pO,linlogMa=pLL),
              mlf(covarMa="DoseN=0", ordinMa="DoseO=0", linlogMa="DoseLL=0"))
							
\end{verbatim}
\normalsize
Although the Tukey-trend test regression models are significant, the pairwise contrasts are as in chapter \ref{green} whereas for pairwise comparisons the regression-type confidence intervals are equivalent to the t-test-type intervals (only here), see Figure \ref{fig:GreenTukey}. The effect sizes are quite different: for the Tukey-type regression models it is the slope parameter, whereas for the pairwise contrasts it is the odds ratios of proportions.

%%%%%%%%%%%%%%%%%%%%%%%%%%%%%%%%%%%%%%%%%%%%%%%%%%%%%
\section{Conclusions}
Evidence of a strictly monotonic trend requires, on the one hand, a common, i.e., monotonic trend statement and, in addition, non-inferiority evidence that the effects of lower doses are not clearly in the opposite direction. I.e. the claim for strict monotonic trend is depending on not well-defined non-inferiority threshold $\eta$ or its post-hoc choice, as proposed by means of a pattern of confidence intervals.
This paper is restricted to a completely randomized one-way layout- however extensions to higher-way layouts or even more complex mixed effect models are possible. Notice, hypotheses trend tests provide not model selection properties, i.e. the particular single contrast with the smallest p-value stands not necessarily for the true model. As a consequence of Paracelsus paradigm, in risk assessment  sometimes a claim for no trend is needed when performing a proof of hazard approach). The above confidence interval approach can be used accordingly. At the end there is a deviation of goods if model assumption and data contradict each other. Should we use the monotonic, even the linear model, regardless of the real data or should we take into account effects against the assumed trend at smaller doses?

%%%%%%%%%%%%%%%%%%%%%%%%%%%%%%%%%%%%%%%%%%%%%%%%%%%%%%%%
\footnotesize
\section*{Appendix}
\subsection*{Designs without control groups}\label{controls}
\begin{figure}[htbp]
	\centering
		\includegraphics[width=0.40\textwidth]{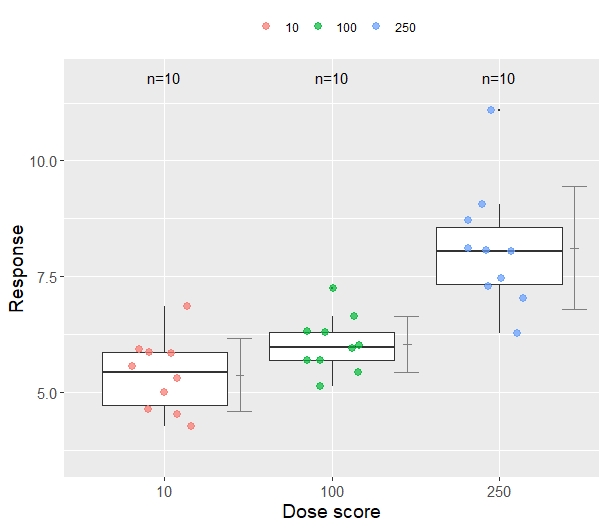}
		\includegraphics[width=0.40\textwidth]{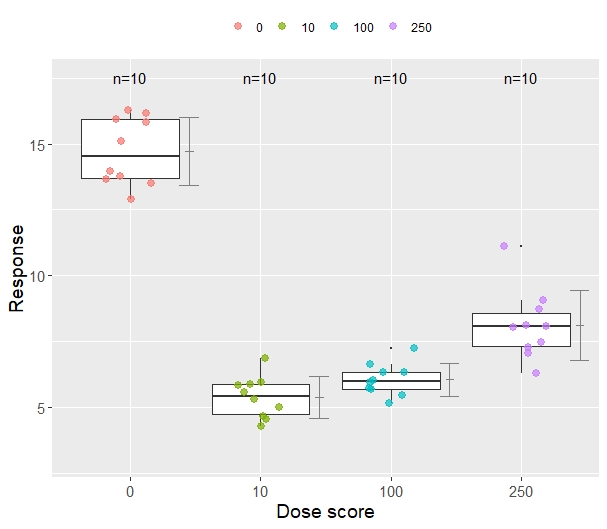}
	\caption{Possible bias when using designs without a negative control}
	\label{fig:without}
\end{figure}

Common dose-response studies use a negative control or placebo group. In principle, designs without a negative control can be used, however they are easily subject to bias, as shown in a teaching example below. In the left panel a strong increasing trend can be claimed ($p_{Tukey trend}<0.0001$, whereas this is not the case when considering the negative control (right panel), see Figure \ref{fig:without}.

A similar bias can occur when not using additionally a positive control, i.e. a treatment with established effect. 
Again using a further  teaching example, the question may arise, what is the value of a clearly increasing trend, but which is absolutely negligible in relation to the proven effect in a C+, i.e. e.g. the comparator in the market?

 \begin{figure}
	 \centering
		 \includegraphics[width=0.4\textwidth]{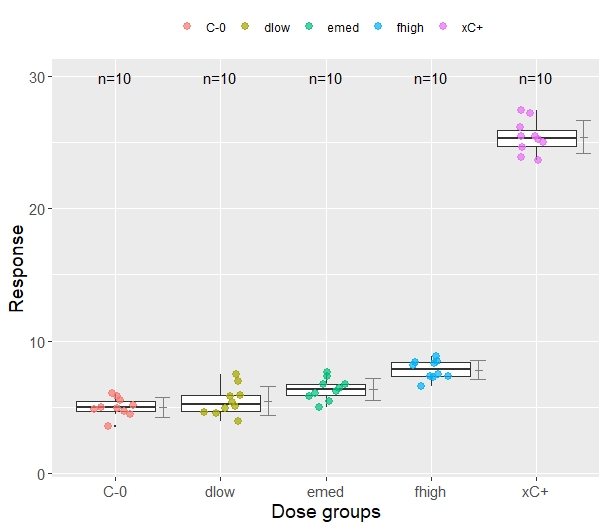}
	 \caption{Dose-response relationship and C+}
	 \label{fig:withP}
 \end{figure}

The consequence of avoiding such distortion is to use both C- and C+ in the design. This can be used to demonstrate both assay sensitivity 
$\mu_{C+}-\mu_{C-}$ and at least non-inferiority of at least one dose with respect to C+ \cite{Bauer1998}.

\subsection*{A proposal for an approach, able to claim a trend where each change point contrast contribute}\label{change}
Change point contrasts are defined to compare specific combined dose levels, in the Shirley data example:

\begin{table}[ht]
\centering\footnotesize
\begin{tabular}{rrrrr}
  \hline
 Contrast& 0 & 1 & 2 & 3 \\ 
  \hline
C 1 & -1.00 & 0.33 & 0.33 & 0.33 \\ 
  C 2 & -0.50 & -0.50 & 0.50 & 0.50 \\ 
  C 3 & -0.33 & -0.33 & -0.33 & 1.00 \\ 
   \hline
\end{tabular}
\caption{Change point contrast matrix}
	\label{tab:matC}

\end{table}

\footnotesize
\begin{verbatim}
change<-summary(glht(mod2, linfct = mcp(Dose="Changepoint"),
         alternative="greater", vcov = sandwich))
\end{verbatim}
\normalsize

Not surprising, in Shirley s example all change points are significant
\begin{table}[ht]
\centering\footnotesize
\begin{tabular}{rlrrrrr}
  \hline
 Contrast& Estimate & SE & test statistics & adjusted p-value \\ 
  \hline
C 1 & 3.5267 & 0.4780 & 7.3781 & 0.0000 \\ 
  C 2 & 3.0100 & 0.6963 & 4.3227 & 0.0001 \\ 
  C 3 &  3.1133 & 0.8576 & 3.6303 & 0.0013 \\ 
   \hline
\end{tabular}
\caption{Reaction time assay data: change point multiple contrast test}
	\label{tab:CTShir}

\end{table}

\subsection*{A proposal for an approach, able to claim a trend where each dose increment contribute}\label{inc}
A specific version of trend is the assumption that each dose increment contribute, i.e. a sequence contrast is used:
\begin{table}[ht]
\centering\footnotesize
\begin{tabular}{rrrrr}
  \hline
 Contrast & 0 & 1 & 2 & 3 \\ 
  \hline
1 - 0 & -1.00 & 1.00 & 0.00 & 0.00 \\ 
  2 - 1 & 0.00 & -1.00 & 1.00 & 0.00 \\ 
  3 - 2 & 0.00 & 0.00 & -1.00 & 1.00 \\ 
   \hline
\end{tabular}
\caption{Contrast matrix for pairwise increments}
	\label{tab:matP}

\end{table}
\footnotesize
\begin{verbatim}
seq<-summary(glht(mod2, linfct = mcp(Dose="Sequen"),
\end{verbatim}
\normalsize

Not all increments contribute, only the increment from $0\Rightarrow 1$.
\begin{table}[ht]
\centering\footnotesize
\begin{tabular}{rlrrrrr}
  \hline
 Contrast & Estimate & SE & test statistics & adjusted p-value \\ 
  \hline
1 - 0 & 2.2800 & 0.7913 & 2.8813 & 0.0093 \\ 
  2 - 1 & 1.0400 & 1.1575 & 0.8984 & 0.5187 \\ 
  3 - 2 & 1.6600 & 1.1460 & 1.4485 & 0.2272 \\ 
   \hline
\end{tabular}
\caption{Reaction time assay data: pairwise increment multiple contrast test}
	\label{tab:MpShir}

\end{table}

\section*{An approach, able to detect an unwanted trend statement} \label{onlykS}
\subsection*{A proposal for an approach, able to detect an unwanted trend statement when supported by $\mu_k$ only} \label{onlyk}
A possible elementary  alternative is $\mu_0=\mu_1=...=\mu_{k-1}<\mu_k$. The both regression models and the Williams  trend test are sensitive for the rather specific alternative. If one does not want to interpret such an alternative as a monotonic trend, a specific test is needed. E.g. a MCT consisting of all pairwise contrasts vs. control (i.e. Dunnett test), where only the kth contrast is under $H_1$, all the other remain under $H_0$. This specific pattern represents such an unwanted trend statement. Notice, due to its pooling properties, the Williams trend test can be be used for this purpose. Simultaneous testing of Dunnett and Williams trend test contrasts is conceivable \cite{Jaki2013}. Here one recognizes strictly monotonic trend but also the unwanted trend simultaneously.
%%%%%%%%%%%%%%%%%%%%%%%%%%%%%%%%%%%%%%%%%
\subsection*{A proposal for an approach, able to detect an unwanted trend statement when supported by a plateau profile only}
A possible elementary  alternative is $\mu_0<\mu_1=...=\mu_{k-1}=\mu_k$, i.e. a plateau profile. While regression models are less powerful, the Williams  trend test is particular powerful because of its pooling properties. If one does not want to interpret such an alternative as a monotonic trend, a specific test is needed. E.g. a Williams-type MCT can be used where the contrast $(\mu_k+\mu_{k-1}+...+\mu_1)/k-\mu_0$ should reveals the smallest adjusted p-value. This specific pattern represents such an unwanted trend statement. Simultaneous testing of Dunnett and Williams trend test contrasts is conceivable \cite{Jaki2013}. Here one recognizes strictly monotonic trend but also both unwanted trends simultaneously.

\section*{Acknowledgment}
I would like to thank Mrs. Goetschi, SCQM Foundation Swiss Clinical Quality Management in Rheumatic Disease Zurich, Switzerland,   for confronting me with the problem of strict monotonic trends.

     \bibliographystyle{apalike}

      \footnotesize
			
    %  \bibliography{Zurich}

\end{document}